\documentclass[aps,preprint]{revtex4}%
\usepackage{amsfonts}
\usepackage{amsmath}
\usepackage{amssymb}
\usepackage{graphicx}%
\begin{document}
\title[Distribution of local density of states]{Distribution of local density of states in superstatistical random matrix theory}
\author{A. Y. Abul-Magd}
\affiliation{Department of Mathematics, Faculty of Science, Zagazig University, Zagazig, Egypt}

\begin{abstract}
We expose an interesting connection between the distribution of local spectral
density of states arising in the theory of disordered systems and the notion
of superstatistics introduced by Beck and Cohen and recently incorporated in
random matrix theory. The latter represents the matrix-element joint
probability density function as an average of the corresponding quantity in
the standard random-matrix theory over a distribution of level densities. We
show that this distribution is in reasonable agreement with the numerical
calculation for a disordered wire, which suggests to use the results of theory
of disordered conductors in estimating the parameter distribution of the
superstatistical random-matrix ensemble.

\end{abstract}
\date[Date text]{\today}
\startpage{1}
\endpage{2}
\maketitle

\section{Introduction}

The formalism of superstatistics (statistics of a statistics), has recently
been proposed by Beck and Cohen \cite{BC} as a possible generalization of
statistical mechanics. Superstatistics arises as weighted averages of ordinary
statistics (the Boltzmann factor) due to fluctuations of one or more intensive
parameter (e.g. the inverse temperature). It considers a non-equilibrium
system as traveling within its phase space which is partitioned into cells.
Within each cell, the system is described by ordinary Maxwell-Boltzmann
statistical mechanics, i.e., its statistical distribution is the canonical one
$e^{-\beta E}$, but $\beta$ varies from cell to cell, with its own probability
density $f(\beta)$. This formalism has been elaborated and applied
successfully to a wide variety of physical problems, e.g., in
\cite{cohen,beck,beckL,salasnich,sattin,reynolds,ivanova,at,bcs,beckT}.

Superstatistics has been applied to model systems with partially chaotic
classical dynamics within the framework of random-matrix theory (RMT) in Ref.
\cite{supst,sust}. It has provides a possible mechanism for the initial stage
of transition of a system out of the state of chaos but fails to reproduce the
Poisson statistics that is believed to describe regular systems. In the
standard RMT \cite{mehta,haake,guhr,stockmann}, a chaotic system is modeled by
an ensemble of random matrices that depends only on the symmetry of the
system. For example, a system of spinless particles, which has a time-reversal
symmetry is represented by a Gaussian orthogonal ensemble (GOE). The joint
distribution of matrix elements of the Hamiltonian $H$ is proportional to
$\exp\left[  -\eta\text{Tr}\left(  H^{\dagger}H\right)  \right]  ,$ where Tr
is trace and $H^{\dagger}$ is the Hermitian conjugate of $H$. The parameter
$\eta$ is related to the square of the mean level density. This distribution
is based on two main assumptions: (i) the matrix elements are independent
identically-distributed random variables, and (ii) their distribution is
invariant under unitary transformations. For most of the physical systems,
however, the phase space is partitioned into regular and chaotic domains.
These systems are known as mixed systems. Attempts to generalize RMT to
describe such mixed systems are numerous; for a review see \cite{guhr}. Most
of these attempts are based on constructing ensembles of random matrices whose
elements are independent but not identically distributed, e.g. in
\cite{rosen,hussein,casati,fyodorov,haake}. Thus, the resulting expressions
are not invariant under base transformation. The superstatistical
generalization follows another route. It keeps base invariance, but violates
matrix-elements independence. The intuitive explanation for using
superstatistics is based on the ansatz that the spectrum of the system under
consideration is partitioned into small cells. Within each cell, the spectrum
is described by an ordinary Gaussian random-matrix ensemble, but ensemble
parameter $\eta$ varies from cell to cell.. One may define the density of
states constituting a single cell as a `local density of states' (LDOS)
proportional to$\sqrt{\eta}$. Superstatistics assumes that LDOS and thus the
parameter $\eta$ is no more a constant parameter as in the original RMT, but
allowed to fluctuate according to a distribution $\widetilde{f}(\eta)$. The
joint matrix-element distribution is represented as an average over
$\exp\left[  -\eta\text{Tr}\left(  H^{\dagger}H\right)  \right]  $ with
respect to the parameter $\eta$. The resulting distribution depends on the
matrix elements in the form of Tr$\left(  H^{\dagger}H\right)  $, which is
base independent as the corresponding distribution in ordinary RMT.

The central question for superstatistics is how to choose the parameter
distribution. In superstatistical thermodynamics, one obtains Tsallis'
statistics \cite{Ts1,Ts2} when the inverse temperature $\beta$ has a $\chi
^{2}$ distribution, but this is not the only possible choice. Beck and Cohen
give several other possible examples of functions which are possible
candidates for $f(\beta)$. Generalized entropies, which are analogous to the
Tsallis entropy, can be defined for these general superstatistics
\cite{abe,souza}. Sattin \cite{sattin} suggested that, lacking any further
information, the most probable realization of $f(\beta)$ will be the one that
maximizes the Shannon entropy. This latter approach was used in \cite{sust} to
estimate the distribution $\widetilde{f}(\eta)$ of the parameter $\eta$ of RMT.

The object of the present paper is to find out whether the local density of
states defined above is related to the local spectral density of states (in
the literature, also LDOS) which is relevant to many practical applications in
the field of condensed matter physics \cite{mirlin}. There, the notion of LDOS
follows from the conjecture that even in a metallic sample there is a finite
probability to find \textquotedblleft almost localized\textquotedblright%
\ eigenstates so that the density of eigenstates is different at different
locations in the sample. Here, by localization we shall mean a situation, in
which eigenfunctions are localized in the space of "unperturbed" eigenvalues
on a scale which is significantly smaller than the size of the ensemble.
Indeed, the size of the region which is populated by an eigenfunction (termed
localization length) measures the maximum number of basis state coupled by
perturbation (off-diagonal elements). In the state of chaos, the localization
length approaches the size of the ensemble, which means that the
eigenfunctions become ergodic, i.e., extended over the whole energy shell.
Then, the LDOS coincides with the "global" density of states and both follow
Wigner's semicircular law. As the system departs from chaos, the localization
length decreases and the LDOS becomes distinguished from the global one. The
main question, which we seek to answer is the following: \textquotedblleft Can
we benefit from the achievements of the well developed theory of disordered
conductors in estimating the parameter distribution $\widetilde{f}(\eta)$ of
the superstatistical ensemble?" Instead of answering this question directly
(which is technically quite difficult), we compare the superstatistical
distribution $\widetilde{f}(\eta)$ derived in \cite{sust} from the principle
of maximum entropy with the distribution of LDOS deduced by Altshuler and
Prigodin \cite{altsher} for disordered conductors.\ After a brief review of
our previous results, given in Section II, we show in Section III that the two
distributions have similar shapes especially near the chaotic limit. The
distribution $\widetilde{f}(\eta)$ is compared with the distribution of LDOS
obtained in the numerical simulation of the closed wire \cite{schomerus}.
While the agreement of the superstatistical distribution with the numerical
experiment is not completely satisfactory, it still demonstrates the analogy
between the notions of LDOS in the two disciplines. Section IV shows by
comparison with a numerical experiment \cite{gu} that the level-density
distribution obtained for a random wire can be used as a parameter
distribution in the superstatistical random matrix theory. The conclusion of
this work is formulated in Section V.

\section{FORMALISM}

For the sake of completeness and clarity, we recall the derivation of the
superstatistical model introduced in \cite{sust}. RMT models the Hamiltonian
of a chaotic system in terms of an ensemble of random matrices, whose matrix
elements have a Gaussian probability density distribution
\begin{equation}
P\left(  H\right)  \varpropto\exp\left[  -\eta\text{Tr}\left(  H^{\dagger
}H\right)  \right]  .
\end{equation}
The mean level density for a GOE with a large dimension $N$ is given by
Wigner's semi-circle law \cite{mehta}%
\begin{equation}
\rho(\eta,\varepsilon)=\frac{1}{2\pi}\left[  N\eta(1-\eta\varepsilon
^{2}/N)\right]  ^{1/2}~\Theta(1-\eta\varepsilon^{2}/N),
\end{equation}
where $\varepsilon$ is the eigenvalue of $H$\ and $\Theta(X)$ is the Heaviside
step function. In practical calculations with GOE, one usually avoids the ends
of the spectrum and works in a region with a nearly constant level density
equal to $\rho(\eta,0)\sim\sqrt{\eta}$.

The superstatistical generalization models the quantum-number space of a
system of mixed regular-chaotic dynamics as made up of many smaller cells that
are temporarily in a chaotic phase. Each cell is large enough to obey the
statistical requirements of RMT but has a different distribution parameter
$\eta$ associated with it, according to a probability density $\widetilde
{f}(\eta)$. Consequently, the superstatistical random-matrix ensemble
describes the mixed system as a mixture of Gaussian ensembles. Its
matrix-element joint probability density distributions obtained by integrating
distributions of the form in Eq. (1) over all positive values of $\eta$\ with
a statistical weight $\widetilde{f}(\eta)$,
\begin{equation}
P(H)=\int_{0}^{\infty}\widetilde{f}(\eta)\frac{\exp\left[  -\eta
\text{Tr}\left(  H^{\dagger}H\right)  \right]  }{Z(\eta)}d\eta,
\end{equation}
where $Z(\eta)=\int\exp\left[  -\eta\text{Tr}\left(  H^{\dagger}H\right)
\right]  d\eta$. Here we use the \textquotedblright B-type
superstatistics\textquotedblright\ \cite{BC}. The parameter $\eta$ may be
expressed in terms of the local mean level spacing $D$\ as%
\begin{equation}
D=\frac{c}{\sqrt{\eta}},
\end{equation}
where $c$ is a constant depending on the size of the ensemble, which can be
evaluated by setting $D=1/\rho(\eta,0)$ and using Eq. (2).

In the new framework of RMT provided by superstatistics, the local mean
spacing $D$ is no longer a fixed parameter but it is a stochastic variable
with probability distribution $f(D)$. Instead, the observed mean level spacing
is just its expectation value. The fluctuation of the local mean spacing is
due to the correlation of the matrix elements which disappears for chaotic
systems. In the absence of these fluctuations, $f(D)=\delta(D-1)$ and we
obtain the standard RMT. Within the superstatistics framework, we can
express\ any statistic $\sigma$ of a (sufficiently chaotic) mixed system that
can in principle be\ obtained from the joint eigenvalue distribution by
integration over some of the eigenvalues, in terms of the corresponding
statistic $\sigma^{(G)}(D)$ for a Gaussian random ensemble with mean level
spacing $D$. The superstatistical generalization is given by
\begin{equation}
\sigma=\int_{0}^{\infty}f(D)\sigma^{(G)}(D)dD.
\end{equation}
The remaining task of superstatistics is the computation of the distribution
$f(D)$. Following Sattin \cite{sattin}, we use the principle of maximum
entropy (MaxEnt) to evaluate the distribution $f(D)$. Lacking a detailed
information about the mechanism causing the deviation from the prediction of
RMT, the most probable realization of $f(D)$ will be the one that extremizes
the Shannon entropy
\begin{equation}
S=-\int_{0}^{\infty}f(D)\ln f(D)dD
\end{equation}
with the following constraints:\textbf{ (}i) The major parameter of RMT is
$\eta$ defined in Eq. (1). Superstatistics was introduced in Eq. (3) by
allowing $\eta$ to fluctuate around a fixed mean value $\left\langle
\eta\right\rangle $. This requires the existence of the mean inverse square of
$D$, $\left\langle D^{-2}\right\rangle =\int_{0}^{\infty}f(D)D^{-2}dD.$ (ii)
The fluctuation properties are usually defined for unfolded spectra, which
have a unit mean level spacing. We thus require $\int_{0}^{\infty}f(D)DdD=1.$
As a result, we obtain
\begin{equation}
f_{\text{MaxEnt}}(D)=C\exp\left[  -\alpha\left(  \frac{2D}{D_{0}}+\frac
{D_{0}^{2}}{D^{2}}\right)  \right]
\end{equation}
where $\alpha$ and $D_{0}$ are parameters, which can be expressed in terms of
the Lagrange multipliers of the constrained extremization, and $C$ is a
normalization constant. Substituting this distribution into Eq. (5) one
obtains expressions for the (global) level density, the
nearest-neighbor-spacing (NNS) distribution and the two-level correlation
function \cite{sust}. Eq. (5) has recently been applied \cite{TrInt} to obtain
a generalization of the well-known Porter-Thomas distribution of transition
intensities, relevant for chaotic regimes, for systems with mixed
regular-chaotic dynamics. This generalization agrees with the data better than
currently available results that can not explain observed shift of the peak
position as the system evolves out of the state of chaos.

\section{DISTRIBUTION OF LOCAL DENSITY OF STATES}

The local density of states plays a central role in the theory of disordered
metals \cite{mirlin,fyodorov1}. It also known in nuclear physics as strength
function \cite{bohr}. It gives the distribution of basis eigenfunctions in
terms of eigenstates of the system. It is obtained by projecting a basis state
$k$ onto exact eigenstates $i$ of the Hamiltonian $H$ and then defined in
terms of the expansion coefficients $C_{k}^{i}$ as%
\begin{equation}
\nu_{k}(E)=\overline{\left\vert C_{k}^{i}\right\vert ^{2}}\rho(E),
\end{equation}
with $\rho(E)$ as the density of exact eigenstates. Here, the average is taken
over a small window of the eigenstates $i$ with energies around $E$. The LDOS
has a well-defined classical interpretation as shown by Benet et al.
\cite{benet}. The unperturbed energy $E_{0}$ is not constant along a classical
trajectory of the full Hamiltonian with a given total energy $H=E$. If we keep
the unperturbed energy $E_{0}$ fixed, the bundle of trajectories of the total
Hamiltonian $H$, which reach the surface of the unperturbed Hamiltonian
$H_{0}=E_{0}$, has a distribution in the total energy $E$ which is described
by a measure $\nu_{E_{0}}(E)$. In the quantum case, this measure corresponds
to the imaginary part of the retarded Green's function at energy $E_{0}$
\cite{casati1}. Distributions of LDOS are relevant for description of
fluctuations of tunneling conductance across a quantum dot \cite{alhassid}, of
transport phenomena in disordered wires \cite{beenakker}, of properties of
some atomic spectra \cite{devries}, and of a shape of NMR line \cite{nmr}. In
random media the LDOS is a random quantity. Altshuler and Prigodin
\cite{altsher} have studied the LDOS distribution in strictly one-dimensional
disordered chains. They obtain closed-form expressions for the LDOS
distributions in open and closed wires. The normalized distribution for the
closed system reads
\begin{equation}
P_{\text{AP}}(\omega,\nu_{\text{AP}};\nu)=\sqrt{\frac{\omega}{2\pi}}\frac
{\nu_{\text{AP}}^{1/2}}{\nu^{3/2}}\exp\left[  \omega-\frac{1}{2}\omega\left(
\frac{\nu}{\nu_{\text{AP}}}+\frac{\nu_{\text{AP}}}{\nu}\right)  \right]  ,
\end{equation}
where $\omega$ is a positive parameter. The distribution $P_{\text{AP}}%
(\omega,\nu_{\text{AP}};\nu)$ has a unit mean value $\nu_{\text{AP}}$ and a
variance equal to $1/\omega$. We note that this distribution fulfils the two
conditions imposed by Beck and Cohen \cite{BC} on the parameter distribution
in\ superstatistics, which are normalization and tending to a delta function
as $\omega\rightarrow\infty$.

Let us now assume that the LDOS $\nu$ is equal to the inverse of local mean
level spacing $D$. In order to justify this assumption, we compare the
distribution $P_{\text{AP}}(\nu)$ with the distribution $f_{\text{MaxEnt}}(D)$
of the local mean level densities (7), which has been applied in the
superstatistical RMT \cite{sust,TrInt}. For this purpose, we have to translate
the distribution $f_{\text{MaxEnt}}(D)$ obtained in the previous section into
the context of the problem of LDOS in complex systems. This can be
straightforwardly done by substituting $\nu=1/D$ in Eq. (7). The distribution
of LDOS in the superstatistical RMT is then given by
\begin{equation}
P_{\text{MaxEnt}}(\nu)=C\frac{1}{\nu^{2}}\exp\left[  -\alpha\left(  \frac
{2\nu_{0}}{\nu}+\frac{\nu^{2}}{\nu_{0}^{2}}\right)  \right]  ,
\end{equation}
where $\nu_{0}=1/D_{0}$. The normalization constant is given by%
\begin{equation}
C=\frac{2\sqrt{\pi}\alpha\nu_{0}}{G_{03}^{30}\left(  \left.  \alpha
^{3}\right\vert 0,\frac{1}{2},1\right)  },
\end{equation}
where $G_{0,3}^{3,0}\left(  z\left\vert b_{1},b_{2},b_{3}\right.  \right)  $
is Meijer's G-function \cite{mathai,wolfram}. The parameter $\nu_{0}$\ is
fixed by requiring that the average value of the LDOS is equal to 1, which
yields%
\begin{equation}
\nu_{0}=\frac{G_{03}^{30}\left(  \left.  \alpha^{3}\right\vert 0,\frac{1}%
{2},1\right)  }{a~G_{03}^{30}\left(  \left.  \alpha^{3}\right\vert
0,0,\frac{1}{2}\right)  }.
\end{equation}
The variance of this distribution is given by%
\begin{equation}
\sigma_{\text{MaxEnt}}^{2}=\frac{G_{03}^{30}\left(  \left.  \alpha
^{3}\right\vert -\frac{1}{2},0,0\right)  G_{03}^{30}\left(  \left.  \alpha
^{3}\right\vert 0,\frac{1}{2},1\right)  }{\left[  G_{03}^{30}\left(  \left.
\alpha^{3}\right\vert 0,0,\frac{1}{2}\right)  \right]  ^{2}}-1,
\end{equation}
which tends to zero as $\alpha$\ tends to $\infty$\ (nearly chaotic regime)
and behaves as $\alpha^{-3/2}$\ at small $\alpha$ (nearly integrable). Figure
1 compares the distributions of LDOS in Eq. (9) and (10) for same variances.
The figure clearly suggests the equivalence of the two distributions are
nearly equivalent, except at small values of $\alpha$ or $\omega$ that
correspond to variances $\sigma^{2}>\left\langle \nu\right\rangle ^{2}$, which
is adapted in the superstatistical approach with nearly integrable systems.
This should not make an obstacle for using the Altshuler-Prigodin LDOS
distribution in superstatistical RMT. The superstatistical approach is not
expected to model a system in the final stage of transition out of chaos,
where the system approaches the integrability limit. As previously mentioned,
this is a base-invariant approach. Integrable systems by definition have well
defined complete set of eigenvalues that yields a diagonal Hamiltonian matrix
when used as a basis.

Schomerus et al \cite{schomerus} calculate the probability distribution of the
local density of states in a disordered one dimensional conductor or
single-mode waveguide. They start from the relation between the LDOS and the
retarded Green function, which they have obtained by a numerical simulation
for a wire with one or both fixed ends . They have given exact results for the
distributions of the local densities of states in one-dimensional
localization, contrasting the microscopic length scale (below the wavelength)
and mesoscopic length scale (between the wavelength and the mean free path).
Their data points are shown in Fig.2 for the case of a closed wire. These data
result from a numerical simulation for a wire of length equal to 55 times the
mean free path, with the LDOS computed halfway in the wire. Figure 2 also
shows the result of calculation using the Altshuler-Prigodin distribution (9)
by a dashed line and the distribution in Eq. (9) by a solid line. The mean
LDOS is fixed to be equal 1 in the three distributions. While the
Altshuler-Prigodin distribution fits the results of numerical simulation
better, the superstatistics distribution still provides a reasonable
representation of the data.

The absence of quantitative agreement between the predictions of Eqs. (9) and
(10) does not imply that the LDOS $\nu$ is not equal to the inverse of local
mean level spacing $D$. The Altshuler-Prigodin distribution (9) is derived
from an elaborate theory that starts from the relation between the level
density and the retarded Green's function for noninteracting electrons in a
wire and arrives at a relation between the microscopic LDOS and the reflection
coefficients, and then uses elaborate methods perform the local spatial
average that gives the LDOS. It is interesting to note that the distribution
in Eq. (10) is derived in \cite{sust} from the principle of maximum entropy
with the constraints of fixed $\left\langle D^{-2}\right\rangle $ and
$\left\langle D\right\rangle $\ as mentioned above. If we replace the first
constraint by another one that requires that $\left\langle D^{-1}\right\rangle
$, we arrive exactly to the Altshuler-Prigodin distribution. The fact that the
Altshuler-Prigodin distribution fits the numerical-experimental data
perfectly, as shown in Fig. 1, while the distribution (10) fits the data only
qualitatively suggests that $f(D)$ may better be obtained by maximizing
entropy under the conditions of fixed $\left\langle D\right\rangle $ and
$\left\langle D^{-1}\right\rangle $.

\section{NEAREST-NEIGHBOR-SPACING DISTRIBUTION}

The question that this section tries to answer is whether the distribution of
LDOS obtained in the study of disordered metals is suitable for describing the
transition out of chaos within the superstatistical approach to RMT. For this
purpose we consider a system, which is described by a superposition of random
matrix ensembles of different LDOS with probability-density function given by
the Altshuler-Prigodin distribution $P_{\text{AP}}(\omega,\nu_{\text{AP}}%
;\nu)$. Its spectral characteristics are expressed as averages of the
corresponding characteristics of a Gaussian random ensemble in analogy with
Eq. (5). Then, the superstatistical NNS distribution is given by%
\begin{equation}
p_{\text{AP}}(\omega,s)=\int_{0}^{\infty}P_{\text{AP}}(\omega,\nu_{\text{AP}%
};\nu)~p_{\text{WD}}(\nu,s)~d\nu,
\end{equation}
where $p_{\text{WD}}(\nu,s)$ is a Wigner-Dyson distribution with mean level
spacing equal to $1/\nu$,%
\begin{equation}
p_{\text{WD}}(\nu,s)=\frac{\pi}{2}\nu^{2}s\exp\left(  -\frac{\pi}{4}\nu
^{2}s^{2}\right)  .
\end{equation}
We then obtain%
\begin{equation}
p(\omega,s)=\frac{1}{4}\sqrt{2\pi\omega}\nu_{\text{AP}}^{2}s\int_{0}^{\infty
}\sqrt{x}~\exp\left[  \omega-\frac{\omega}{2}\left(  x+\frac{1}{x}\right)
-\frac{\pi}{4}\nu_{\text{AP}}^{2}x^{2}s^{2}\right]  dx
\end{equation}
the parameter $\nu_{\text{AP}}$ is fixed bey requiring that the mean-level
spacing $\left\langle 1/\nu\right\rangle $ is equal to 1 which yields
\begin{equation}
\nu_{\text{AP}}=1+1/\omega.
\end{equation}
Unfortunately, we are not able to evaluate the integral in Eq. (16)
analytically, but there is no problem in its numerical evaluation.

We shall demonstrate the quality of the NNS distribution in Eq. (16) by
applying it to the results of a numerical experiment by Gu et al. \cite{gu} on
a random binary network. Impurity bonds are employed to replace the bonds in
an otherwise homogeneous network. The authors of Ref. \cite{gu} numerically
calculated more than 700 resonances for each sample. For each impurity
concentration $p$, they considered 1000 samples with totally more than 700 000
levels computed. Figure 3 shows their results for four values of concentration
$p$. The figure also shows the best fits obtained for NNS superstatistical
distributions obtained using the parameter distributions following the
Altshuler-Prigodin formula and the ones obtained in \cite{sust} using the
principle of MaxEnt. The high statistical significance of the data allows us
to assume the advantage of the superstatistical distribution base on the
Altshuler-Prigodin LDOS distribution for describing the results of this experiment.

\section{SUMMARY AND CONCLUSION}

Many characteristics of disordered metals and those of insulators have been
understood using RMT. The spectral fluctuations of a disordered metal are well
described by random matrix theory, while the fluctuations of a system in the
insulating regime follow the Poisson statistics. The nature and the details of
the metal-insulator transition, on the other hand, still belong to the most
intensively studied problems. Localization of wave functions by disorder can
be seen in the fluctuations of the density of states, which can be probed
using the tunnel resistance of a point contact. We have previously described
the analogous transition between regular and chaotic dynamics within the
framework of RMT as a superposition of two statistics, namely one described by
the matrix-element distribution $\exp\left[  -\eta\text{Tr}\left(  H^{\dagger
}H\right)  \right]  $\ and another one by the probability distribution
$\widetilde{f}(\eta)$ of the inverse variance of the matrix elements, which is
derived from the principle of maximum entropy. In the present investigation,
we show an interesting connection between the distribution $\widetilde{f}%
(\eta)$ and the distribution of LDOS which is used in the study of disordered
systems. We have also found that the LDOS distribution that follows from
$\widetilde{f}(\eta)$ is in a reasonable agreement with the numerical
numerical simulation of LDOS is a weakly-disordered metallic wire. The
agreement is of course not as good as for the distributions obtained by
Altshuler and Prigodin for LDOS in one-dimensional disordered systems using a
more sophisticated approach. However,we obtain the Altshuler-Prigodin
distribution if we modify the (optional) constraints imposed on entropy
maximization in our previous work, by requiring a fixed value of $\left\langle
\eta^{1/2}\right\rangle $ rather than fixing $\left\langle \eta\right\rangle
$. Therefore, both the Altshuler-Prigodin distribution and the parameter
distribution used in our previous work are equally well based on the principle
of maximum entropy, both satisfy the Beck-Cohen criteria, but the comparison
with experimental data prefers the former distribution. In conclusion, we
suggest to use the Altshuler-Prigodin formula as a parameter distribution that
describes the superposition of GOE's in the superstatistical ensemble. This
has been illustrated by an analysis of a high-quality numerical experiment on
the NNS distributions of resonance spectra of disordered binary networks.

{\LARGE Figure captions}

FIG. 1. (Color on line) Distributions of LDOS estimated by using the principle
of MaxEnt and previously used in the superstatistical approach to RMT
\cite{sust} (solid line) compared with the ones obtained by Altshuler and
Prigodin for a closed one-dimensional wire \cite{altsher} (dashed line) that
have same variances.

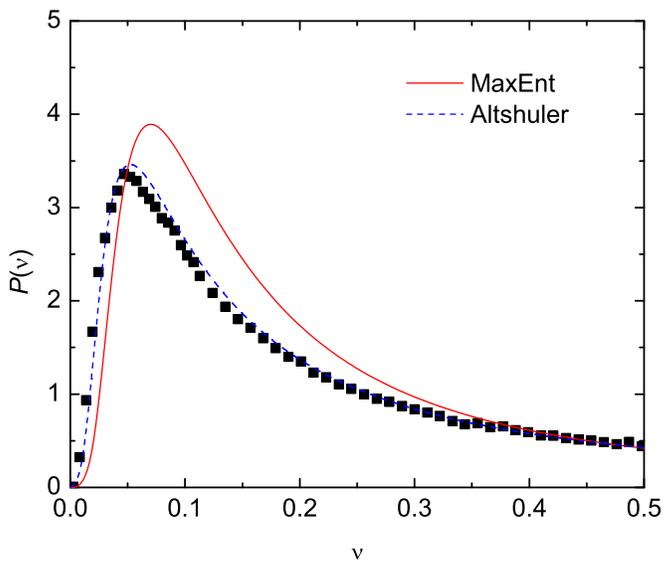
FIG. 2. ((Color on line) Distribution of LDOS in a disordered metal
numerically calculated by Schomerus et al. \cite{schomerus} (histogram)
compared with the corresponding distribution deduced from the superstatistical
approach to RMT \cite{sust} (solid line) and that obtained by Altshuler and
Prigodin for a closed one-dimensional wire \cite{altsher} (dashed line).

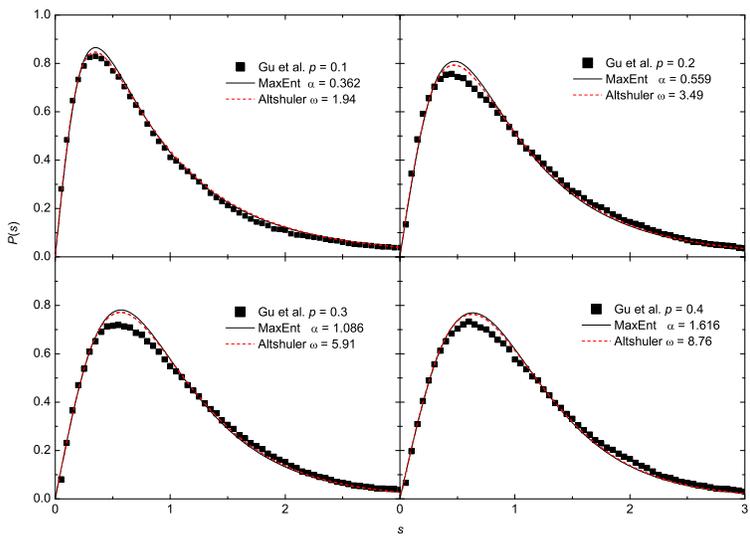
FIG. 3. (Color on line) Nearest neighbor spacing distributions of geometrical
resonances in random network, calculated by Gu et al. \cite{gu} compared with
the superstatistical distributions in which the parameter distributions are
estimated using the MaxEnt principle (solid lines) as well as those based on
the Altshuler-Prigodin distribution of LDOS (dashed lines).

\pagebreak
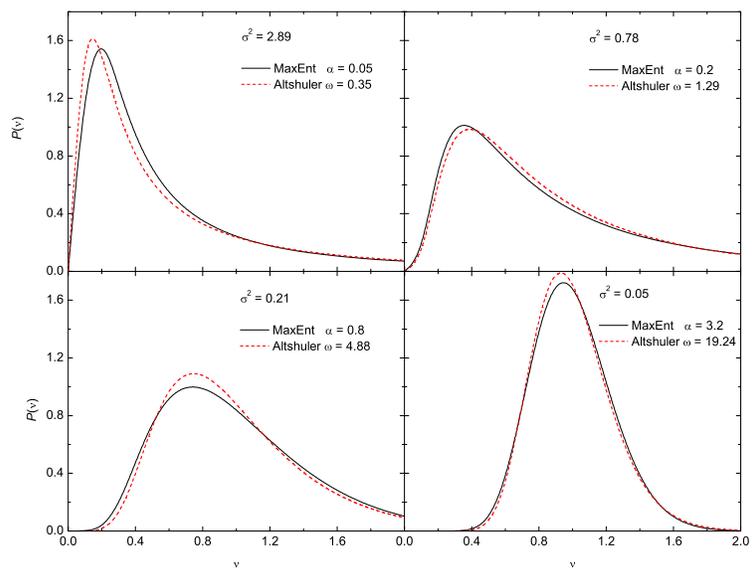
\end{document}